\documentclass[aps,prx,twocolumn,amsmath,amssymb,showpacs, superscriptaddress,notitlepage,longbibliography,10pt]{revtex4-2}

\usepackage[colorlinks=true,linkcolor=blue,anchorcolor=red,citecolor=blue, urlcolor=blue]{hyperref}
\usepackage{bm}
\usepackage{graphicx}
\usepackage{color}
\usepackage[dvipsnames]{xcolor}
\usepackage{multirow}
\usepackage{extarrows}
\usepackage{array}
\usepackage{appendix}
\usepackage{xr}
\usepackage{bbding}
\usepackage{cases}
\usepackage[normalem]{ulem}

\usepackage{tikz}

\definecolor{iconSKFill}{RGB}{200,227,247}
\definecolor{iconSJFill}{RGB}{223,225,227}
\definecolor{iconLine}{RGB}{150,166,180}
\definecolor{iconSJLine}{RGB}{128,136,140}
\definecolor{iconGeoFill}{RGB}{199,224,185} 
\definecolor{iconGeoLine}{RGB}{132,156,124}

\newcommand{\iconTikZPicture}[2]{%
  \tikz[baseline=0.18em,x=1em,y=1em,scale=#1,line join=round,line cap=round]{#2}%
}

\newcommand{\iconTikZText}[2]{%
  \leavevmode\raisebox{0pt}[0pt][0pt]{\smash{\iconTikZPicture{#1}{#2}}}%
}

\newcommand{\iconTikZMath}[2]{%
  \smash{\vcenter{\hbox{\iconTikZPicture{#1}{#2}}}}%
}

\newcommand{\iconTikZ}[1]{%
  \ifmmode
    \mathchoice
      {\iconTikZMath{1.00}{#1}}%
      {\iconTikZMath{1.00}{#1}}%
      {\iconTikZMath{0.70}{#1}}%
      {\iconTikZMath{0.56}{#1}}%
  \else
    \iconTikZText{1.00}{#1}%
  \fi
}

\newcommand{\iconSquareAt}[1]{%
  \begin{scope}[shift={#1}]
    \draw[fill=iconSKFill,draw=iconLine,line width=0.45pt] (0,0) rectangle (0.72,0.72);
  \end{scope}%
}

\newcommand{\iconTriangleAt}[1]{%
  \begin{scope}[shift={#1}]
    \draw[fill=iconSJFill,draw=iconSJLine,line width=0.62pt] (0,0) -- (0.78,0) -- (0.39,0.72) -- cycle;
  \end{scope}%
}

\newcommand{\iconBerryAt}[1]{%
  \begin{scope}[shift={#1}]
    \path[use as bounding box] (-0.04,-0.04) rectangle (0.80,0.80);
    \begin{scope}
      \clip (0.38,0.38) circle (0.34);
      \foreach \x in {-0.25,-0.08,0.09,0.26,0.43,0.60,0.77,0.94}{
        \draw[iconGeoLine,line width=0.72pt] (\x,-0.10) -- (\x-0.66,0.92);
      }
    \end{scope}
    \draw[draw=iconGeoLine,line width=0.72pt] (0.38,0.38) circle (0.34);
  \end{scope}%
}

\newcommand{\iconMetricAt}[1]{%
  \begin{scope}[shift={#1}]
    \path[use as bounding box] (-0.04,-0.04) rectangle (0.80,0.80);
    \draw[fill=iconGeoFill,draw=iconGeoLine,line width=0.72pt] (0.38,0.38) circle (0.34);
  \end{scope}%
}

\newcommand{\iconTOIAt}[1]{%
  \begin{scope}[shift={#1}]
    \path[use as bounding box] (-0.04,-0.04) rectangle (1.12,0.80);
    \draw[fill=iconGeoFill,draw=iconGeoLine,line width=0.72pt] (0.38,0.38) circle (0.34);
    \begin{scope}
      \clip (0.70,0.38) circle (0.34);
      \foreach \x in {0.07,0.24,0.41,0.58,0.75,0.92,1.09,1.26}{
        \draw[iconGeoLine,line width=0.72pt] (\x,-0.10) -- (\x-0.66,0.92);
      }
    \end{scope}
    \draw[draw=iconGeoLine,line width=0.72pt] (0.70,0.38) circle (0.34);
  \end{scope}%
}

\DeclareRobustCommand{\iconSK}{\iconTikZ{\iconSquareAt{(0,0)}}}
\DeclareRobustCommand{\iconTwoSK}{\iconTikZ{\iconSquareAt{(0,0)}\iconSquareAt{(0.18,0.18)}}}
\DeclareRobustCommand{\iconThreeSK}{\iconTikZ{\iconSquareAt{(0,0)}\iconSquareAt{(0.18,0.18)}\iconSquareAt{(0.36,0.36)}}}

\DeclareRobustCommand{\iconSJ}{\iconTikZ{\iconTriangleAt{(0,0)}}}
\DeclareRobustCommand{\iconTwoSJ}{\iconTikZ{\iconTriangleAt{(0,0)}\iconTriangleAt{(0.18,0.18)}}}
\DeclareRobustCommand{\iconThreeSJ}{\iconTikZ{\iconTriangleAt{(0,0)}\iconTriangleAt{(0.18,0.18)}\iconTriangleAt{(0.36,0.36)}}}

\DeclareRobustCommand{\iconSSK}{\iconTikZ{\iconSquareAt{(0,0)}\iconTriangleAt{(0.24,0.24)}}}
\DeclareRobustCommand{\iconSTwoSK}{\iconTikZ{\iconSquareAt{(0,0)}\iconSquareAt{(0.24,0.24)}\iconTriangleAt{(0.48,0.48)}}}
\DeclareRobustCommand{\iconTwoSSK}{\iconTikZ{\iconSquareAt{(0,0)}\iconTriangleAt{(0.24,0.24)}\iconTriangleAt{(0.48,0.48)}}}

\DeclareRobustCommand{\iconBCQ}{\iconTikZ{\iconBerryAt{(0,0)}}}
\DeclareRobustCommand{\iconQMQ}{\iconTikZ{\iconMetricAt{(0,0)}}}
\DeclareRobustCommand{\iconTOI}{\iconTikZ{\iconTOIAt{(0,0)}}}

\DeclareRobustCommand{\iconBSJ}{\iconTikZ{\iconBerryAt{(0,0)}\iconTriangleAt{(0.24,0.24)}}}
\DeclareRobustCommand{\iconBSK}{\iconTikZ{\iconBerryAt{(0,0)}\iconSquareAt{(0.24,0.24)}}}
\DeclareRobustCommand{\iconBTwoSJ}{\iconTikZ{\iconBerryAt{(0,0)}\iconTriangleAt{(0.24,0.24)}\iconTriangleAt{(0.48,0.48)}}}
\DeclareRobustCommand{\iconBTwoSK}{\iconTikZ{\iconBerryAt{(0,0)}\iconSquareAt{(0.24,0.24)}\iconSquareAt{(0.48,0.48)}}}
\DeclareRobustCommand{\iconBSSK}{\iconTikZ{\iconBerryAt{(0,0)}\iconSquareAt{(0.24,0.24)}\iconTriangleAt{(0.48,0.48)}}}

\DeclareRobustCommand{\iconQSJ}{\iconTikZ{\iconMetricAt{(0,0)}\iconTriangleAt{(0.24,0.24)}}}
\DeclareRobustCommand{\iconQSK}{\iconTikZ{\iconMetricAt{(0,0)}\iconSquareAt{(0.24,0.24)}}}

\begin{document}

\title{Identifying Geometric Third-Order Nonlinear Transport in Disordered Materials}
\author{Zhen-Hao Gong}
\affiliation{State Key Laboratory of Quantum Functional Materials, Department of Physics, and Guangdong Basic Research Center of Excellence for Quantum Science, Southern University of Science and Technology (SUSTech), Shenzhen 518055, China}
\affiliation{Guangdong Provincial Key Laboratory of Topological Matter, Southern University of Science and Technology (SUSTech), Shenzhen 518055, China}

\author{Zhi-Hao Wei}
\affiliation{State Key Laboratory of Quantum Functional Materials, Department of Physics, and Guangdong Basic Research Center of Excellence for Quantum Science, Southern University of Science and Technology (SUSTech), Shenzhen 518055, China}
\affiliation{Guangdong Provincial Key Laboratory of Topological Matter, Southern University of Science and Technology (SUSTech), Shenzhen 518055, China}

\author{Hai-Zhou Lu}
\email{Corresponding author: luhz@sustech.edu.cn}
\affiliation{State Key Laboratory of Quantum Functional Materials, Department of Physics, and Guangdong Basic Research Center of Excellence for Quantum Science, Southern University of Science and Technology (SUSTech), Shenzhen 518055, China}
\affiliation{Guangdong Provincial Key Laboratory of Topological Matter, Southern University of Science and Technology (SUSTech), Shenzhen 518055, China}
\affiliation{Quantum Science Center of Guangdong-Hong Kong-Macao Greater Bay Area (Guangdong), Shenzhen 518045, China}

\author{X. C. Xie}
\affiliation{International Center for Quantum Materials, School of Physics, Peking University, Beijing 100871, China}
\affiliation{Interdisciplinary Center for Theoretical Physics and Information
Sciences (ICTPIS), Fudan University, Shanghai 200433, China}
\affiliation{Hefei National Laboratory, Hefei 230088, China}

\date{\today }

\begin{abstract}
 In nonlinear transport, the quantum-geometric effects can generate higher-harmonic voltages in response to a driving current, which has defined a fast-moving field of intense interest. However, in realistic materials where disorder scattering also contributes to nonlinear transport, identifying the geometric mechanisms remains a challenge. In particular, a theoretical framework for  data analysis is still lacking for nonlinear transport at any order. 
Here, we develop a mechanism-resolved and symmetry-guided framework for identifying mechanisms of third-order nonlinear transport in disordered materials. 
We find a total of 20 mechanisms of third-order nonlinear transport, by treating quantum-geometric and disorder-mediated mechanisms on an equal footing.
More importantly, we propose a protocol of data analysis that combines symmetry diagnosis of magnetic point groups and scaling law of relation between the third-order nonlinear Hall conductivity and linear longitudinal conductivity. We identify characteristic fingerprints in the scaling-law weights, which allow the mechanisms to be quantitatively distinguished in experiments. We have applied the protocol to identify the geometric mechanisms in materials with and without time-reversal symmetry, including 2D materials, topological materials, and altermagnets. The theory can be generalized to arbitrary orders of nonlinear transport, further promoting nonlinear transport as a probe of geometric effects and phase transitions in quantum materials.
\end{abstract}

\maketitle

\section{Introduction}

\begin{figure}[htbp]
\centering
\includegraphics[width=0.33\textwidth]{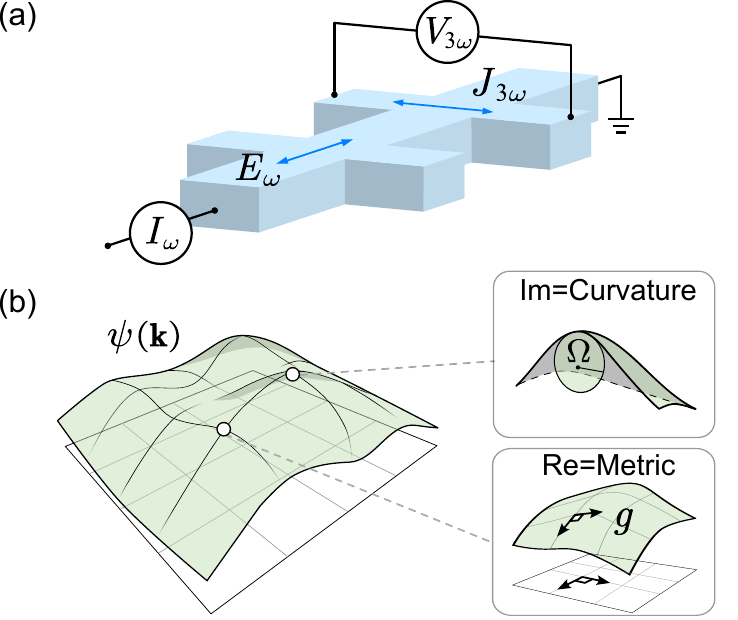}
\caption{(a) Third-order nonlinear Hall effect is measured experimentally as a third-harmonic transverse voltage $V_{3\omega}$ driven by a low-frequency current $I_\omega$ ($\omega \in 10\text{--}1000~\mathrm{Hz}$) or calculated theoretically as a current density $J_{3\omega}$ driven by an electric field $E_\omega$. 
{(b) Schematic illustration of the wave function ($\psi$) manifold in momentum ($\mathbf{k}$) space \cite{crane}; the Berry curvature and quantum metric are the imaginary (Im) and real (Re) parts of the quantum-geometric tensor, respectively. }
}
\label{Fig:QuantumGeometry}
\end{figure}

In standard lock-in measurements of electronic transport, a voltage is measured at the same frequency as a low-frequency ac driving current to filter noise. Instead, the voltage can be measured at integer multiples of that frequency [Fig.~\ref{Fig:QuantumGeometry}(a)], opening a new avenue to explore geometric properties in condensed matter systems, such as the quantum metric dipole \cite{GaoY14prl, Xiao2021PRL,Yang2021PRL,WangJ24prb,YangSY24arXiv,Schnyder25arXiv,LuHZ25arXiv,Nagaosa22prb,Dimi23prb,Kaplan24prl,Wang24prb} and Berry curvature dipole \cite{Fu2015PRL,Low15prb,Du2018PRL} in the second-order nonlinear Hall effect \cite{Ma2019Nature,Kang2019NatMat,Du2021NRP,Xu2023Science,Yan2023Nature}. 
The quantum metric and Berry curvature [Fig.~\ref{Fig:QuantumGeometry}(b)] are, respectively, the real and imaginary parts of the quantum-geometric tensor \cite{Provost80cmp,Resta11epjb}, which measures distances between quantum states and, thus, can provide spacetime insights into the condensed matter problems \cite{Liu24nsr,Bernevig25arXiv,Balazs23pra,Bouhon23arXiv,HuJM24arXiv}, such as in the flat-band superconductivity and fractional Chern insulator \cite{parameswaran2012, roy2014,jackson2015,peotta2015, Torma22nrp,Torma23prl}.
Recently, the third-order nonlinear Hall effect, that is, a transverse voltage at the third harmonic in response to an ac current, has been proposed as a probe of the Berry curvature quadrupole \cite{KTLow23prb}, the quantum metric quadrupole \cite{Yang22prb}, and a third-order intrinsic mechanism that mixes the Berry curvature and quantum metric (a similar term also appears in Einstein's field equation) \cite{Fang24prl,WangJian23prb}, 
attracting tremendous interest in third-order nonlinear transport \cite{Lai21nn,HeP24nn,Wang22nsr,Yu25ncomms,Chen24prb,Shi25prb,Ye22prb,Li24acsami,Li24ncomms,Yang25arxiv,Wei22prb,Nag23prb,Mandal24prb,Pal24prb,Barman25prb}. However, in realistic materials, disorder scattering inevitably coexists  
with the geometric mechanisms \cite{Nagaosa10rmp,Xiao10rmp}. Without comparing the geometric and disorder effects side by side, it is impossible to distinguish the actual mechanisms in experiments.

In this work, we treat geometric effects and disorder scattering on an equal footing and develop a comprehensive theory for the mechanisms of third-order nonlinear transport, including both the longitudinal and Hall components. More importantly, to identify the mechanisms in experiments, we derive a scaling law that expresses the third-order Hall conductivity as a polynomial in the linear longitudinal conductivity. The scaling law reveals distinct weight distributions of the polynomial terms that serve as fingerprints for identifying the dominant geometric and disorder-scattering mechanisms in materials with and without time-reversal symmetry, including 2D materials, topological materials, and unconventional magnets such as altermagnets. Our work further establishes nonlinear transport as a quantitative probe of geometric effects and phase transitions and is broadly applicable to a variety of quantum materials.
\\[2.5em]

\section{Mechanisms of Third-Order Nonlinear Transport}

\begin{figure}[htbp]
\centering
\includegraphics[width=0.43\textwidth]{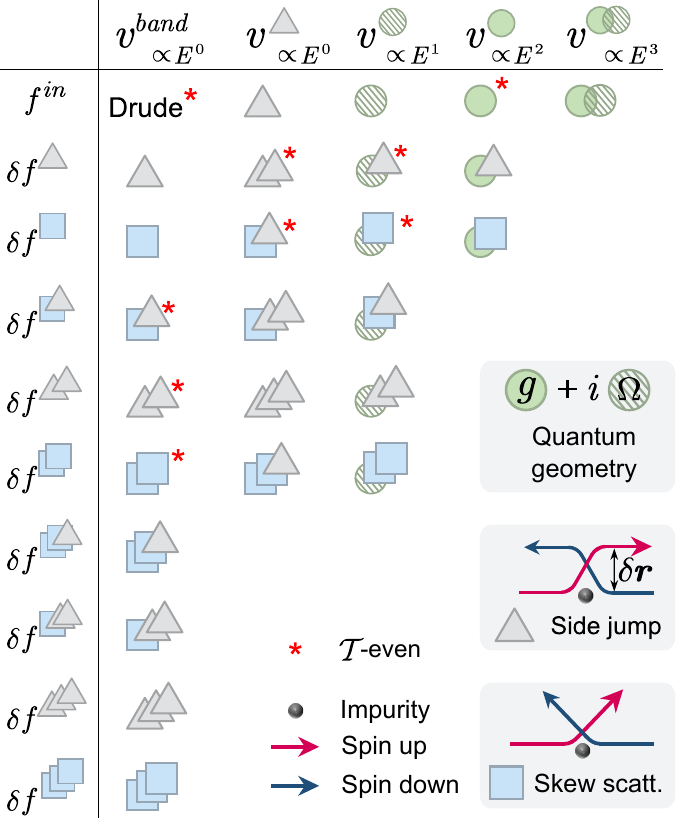}
\caption{We use symbols to visualize the mechanisms of third-order nonlinear transport, which consist of three major elements: the quantum geometry (\iconQMQ + $i$ \iconBCQ), side jump (\iconSJ), which shifts electron coordinates sideways by $\delta\mathbf{r}$, and skew scattering ($\iconSK$), which asymmetrically deflects electrons. 
The horizontal and vertical axes indicate how these elements contribute to the components of the velocity $\dot{\mathbf{r}}$ (and their dependence on the electric field $\mathbf{E}$) and to the nonequilibrium distribution function $\delta f$, respectively. Here, $v^\mathrm{band}$
is the group velocity defined by the band dispersion,
and $f^{in}$ is the distribution function in the relaxation time approximation \cite{Mahan1990}.
The mechanisms of the third-order nonlinear transport are from the combinations of $\dot{\mathbf{r}}$ and $\delta f$ up to third order of $\mathbf{E}$. Besides the known Berry curvature quadrupole (\iconBCQ) \cite{KTLow23prb}, quantum metric quadrupole (\iconQMQ) \cite{Yang22prb}, and third-order intrinsic (\iconTOI) mechanisms \cite{Fang24prl,WangJian23prb}, we discover a total of 20 mechanisms, including 17 more mechanisms related to the disorder scattering, such as the third-order skew scattering (\iconThreeSK). Therefore, how to distinguish the geometric mechanisms in realistic materials, where disorder scattering is ubiquitous, remains a challenge. Mechanisms marked with a red asterisk do not require broken time-reversal symmetry ($\mathcal{T}$-even), and mechanisms without a red asterisk require broken time-reversal symmetry ($\mathcal{T}$-odd); see Sec.~SIIA in Supplemental Material \cite{supp} for the analysis of time-reversal symmetry.}
\label{Fig:Mechanism}
\end{figure}

The third-order nonlinear transport can be derived within the Boltzmann equation formalism. We start from the current density $\mathbf{J}$, which is the sum of the velocity 
$\dot{\mathbf{r}}$ weighted by the nonequilibrium electron distribution $\delta f$.
Both electric field $\mathbf{E}$ and disorder scattering generate different components of $\delta f$, as illustrated on the vertical axis in Fig.~\ref{Fig:Mechanism}, for different orders (e.g., $\delta f^{\iconSK}$ and $\delta f^{\iconThreeSK}$ for the first-order and third-order skew scattering, respectively) and mixed terms (e.g., $\delta f^{\iconSTwoSK}$ for second-order skew scattering combined with first-order side jump). The symbol \iconSJ ~denotes the side jump, which shifts electron coordinates sideways, and \iconSK ~denotes the skew scattering, which deflects electrons asymmetrically. The velocity $\dot{\mathbf{r}}$ contains components from the band group velocity, Berry curvature \cite{Xiao10rmp}, quantum metric \cite{GaoY14prl}, side-jump velocity \cite{Nagaosa10rmp}, and third-order intrinsic velocity that mixes Berry curvature and quantum metric \cite{Fang24prl,Mandal24prb, WangJian23prb}. These components, along with their different dependencies on the electric field $\mathbf{E}$, are summarized on the horizontal axis in Fig.~\ref{Fig:Mechanism}. 
The distribution $\delta f$ can likewise be expanded as a polynomial in $\mathbf{E}$. 
By matching $\dot{\mathbf{r}}$ and $\delta f$ at cubic order in $\mathbf{E}$,
we find a total of 20 mechanisms for the third-order nonlinear conductivity $\chi_{a;bcd}$, defined as 
\begin{eqnarray}\label{Eq:definition-chi}
J_a \equiv \chi_{a;bcd}E_b E_c E_d, 
\end{eqnarray}
where $a,b,c,d\in \{x,y,z\}$ and $J_a$ and $E_{b,c,d}$ are components of $\mathbf{J}$ and $\mathbf{E}$, respectively. 
We visualize 
all the mechanisms in terms of the components of $\delta f$ and $\dot{\mathbf{r}}$ in Fig.~\ref{Fig:Mechanism}. Besides the known geometric mechanisms such as the Berry curvature quadrupole (\iconBCQ) \cite{KTLow23prb}, 
quantum metric quadrupole (\iconQMQ) \cite{Yang22prb}, 
and third-order intrinsic mechanism (\iconTOI) 
\cite{Fang24prl,WangJian23prb},
the new findings here are the distinct mechanisms related to the disorder scattering, such as the third-order skew-scattering (\iconThreeSK) and third-order side-jump (\iconThreeSJ) mechanisms, as well as mixed mechanisms of the geometric effects and disorder scattering, such as \iconBSSK. 
See Sec.~\ref{Sec:calculation-conductivity} for the calculation procedure of the third-order nonlinear conductivity. 
\section{Scaling Law of Third-Order Nonlinear Transport}

Because of the disorder mechanisms discovered in this work, distinguishing geometric mechanisms in experiments is challenging. To address this,  
we derive a scaling law that takes all 20 mechanisms into account. The scaling law expresses the third-order nonlinear Hall conductivity $\chi_{y;xxx}$ as a polynomial in the linear longitudinal conductivity $\sigma_{xx}$. For the $\mathcal{T}$-odd mechanisms, there can be up to seven polynomial terms:
\begin{eqnarray}\label{Eq:Scaling-odd}
\chi_{y;xxx}= \sum_{n=0}^6 \mathcal{C}_n \sigma_{xx}^n,
\end{eqnarray}
and for the $\mathcal{T}$-even mechanisms, there can be up to five terms:
\begin{eqnarray}\label{Eq:Scaling-even}
\chi_{y;xxx}= \sum_{n=1}^5 \mathcal{C}_n \sigma_{xx}^n,
\end{eqnarray}
where the parameters $\mathcal{C}_n$ are associated with distinct mechanisms. By fitting the experimental data of $\chi_{y;xxx}$ as a polynomial function of $\sigma_{xx}$, one can identify which mechanisms dominate in a given experiment.

Specifically, we find that 12 out of 20 mechanisms have unique weight distribution of the polynomial terms in the scaling law, as summarized in Table~\ref{Tab:Weights} 
for all 20 mechanisms of the third-order
nonlinear Hall effect, so the 12 mechanisms can be identified when they dominate in experiments. The remaining eight mechanisms come in twofold degeneracies. For example, $\iconTwoSJ$ and $\iconBSK$  share the ratios $0 : 0 : 0 : 1 : -2 : 1 : 0$. The derivation of the weight ratios follows from the fact that both the linear longitudinal conductivity and nonlinear Hall conductivity depend on disorder scattering. 
For example, the three geometric mechanisms, including the third-order intrinsic mechanism ($\iconTOI$) 
\cite{Fang24prl,WangJian23prb}, the quantum metric quadrupole ($\iconQMQ$) \cite{Yang22prb}, and the Berry curvature quadrupole ($\iconBCQ$) \cite{KTLow23prb}, are characterized by the dominant $\mathcal{C}_0$, $\mathcal{C}_1\sigma_{xx}$, and $\mathcal{C}_2\sigma^{2}_{xx}$ terms, respectively, as shown in the first three rows in Table~\ref{Tab:Weights}. 
This is because the linear longitudinal conductivity $\sigma_{xx}$ is linearly proportional to the scattering time $\tau$ in the semiclassical regime \cite{Ashcroft1976}, and $\iconTOI \propto\tau^0$, $\iconQMQ \propto\tau^1$, and $\iconBCQ  \propto\tau^2$, respectively. Although these geometric mechanisms themselves are known, our key result is that they can be identified experimentally via the scaling law. For the disorder-related mechanisms, the weight ratios are more sophisticated. For instance, the mixed Berry-curvature-plus-side-jump ($\iconBSJ$) mechanism has the weight ratio $\mathcal{C}_2 \sigma_{xx}^2:\mathcal{C}_1 \sigma_{xx} = 1:-1$, because its nonlinear conductivity is proportional to $\tau^2 \langle V_{ll'} V_{l'l}\rangle$, where $\tau^2\propto \sigma^2_{xx}$ arises from the distribution function $\delta f^{\iconSJ}$ and $\langle V_{ll'} V_{l'l}\rangle$ is the correlation of two events of the side-jump scatterings $V_{ll'}$. \begin{table}[thbp]
\caption{Weights of the polynomial terms in the scaling laws for the $\mathcal{T}$-odd [Eq.~(\ref{Eq:Scaling-odd})] and $\mathcal{T}$-even [Eq.~(\ref{Eq:Scaling-even})] mechanisms of the third-order nonlinear Hall effect in Fig.~\ref{Fig:Mechanism}. We find 12 out of the total 20 mechanisms can be identified through the weight distributions, if they are dominant in experiments, such as the third-order intrinsic \iconTOI or third-order skew scattering \iconThreeSK. The remaining eight mechanisms have degenerate weight distributions, such as the second-order side-jump mechanism \iconTwoSJ and the mixed Berry-curvature-plus-skew-scattering mechanism \iconBSK.  }
\label{Tab:Weights}
\setlength{\tabcolsep}{0.05cm}
\renewcommand\arraystretch{1.2}
\begin{tabular}{cccccccccccccc}
\hline
\hline
\textbf{Mech.}&$\mathcal{C}_6\sigma_{xx}^6$&:&$\mathcal{C}_5\sigma_{xx}^5$&:&$\mathcal{C}_4\sigma_{xx}^4$&:&$\mathcal{C}_3\sigma_{xx}^3$&:&$\mathcal{C}_2\sigma_{xx}^2$&:&$\mathcal{C}_1\sigma_{xx}$&:&$\mathcal{C}_0$  
\\
\hline 
$\mathcal{T}$\textbf{-odd}&   & & & & & & & & & & & & 
\\
\iconTOI &   $0$&:&$0$&:&$0$&:&$0$&:&$0$&:&$0$&:&$1$   
\\
\iconQSJ &   $0$&:&$0$&:&$0$&:&$0$&:&$0$&:&$1$&:&$-1$   
\\
\iconBCQ  &  $0$&:&$0$&:&$0$&:&$0$&:&$1$&:&$0$&:&$0$    
\\
\iconSJ &   $0$&:&$0$&:&$0$&:&$1$&:&$-1$&:&$0$&:&$0$   
\\
\iconSK &   $0$&:&$0$&:&$1$&:&$-2$&:&$1$&:&$0$&:&$0$   
\\
\iconSTwoSK &   $0$&:&$1$&:&$-5$&:&$10$&:&$-10$&:&$5$&:&$-1$   
\\
\iconThreeSK &   $1$&:&$-6$&:&$15$&:&$-20$&:&$15$&:&$-6$&:&$1$   
\\
\iconTwoSSK, \iconBTwoSK &   $0$&:&$0$&:&$1$&:&$-4$&:&$6$&:&$-4$&:&$1$  \\
\iconBTwoSJ, \iconQSK &   $0$&:&$0$&:&$0$&:&$0$&:&$1$&:&$-2$&:&$1$   
\\
\iconBSSK, \iconThreeSJ &   $0$&:&$0$&:&$0$&:&$1$&:&$-3$&:&$3$&:&$-1$   
\\
\hline 
$\mathcal{T}$\textbf{-even}&   & & & & & & & & & & & & 
\\
\iconQMQ  &  $0$&:&$0$&:&$0$&:&$0$&:&$0$&:&$1$&:&$0$    
\\
\iconBSJ &   $0$&:&$0$&:&$0$&:&$0$&:&$1$&:&$-1$&:&$0$   
\\
Drude  & $0$&:&$0$&:&$0$&:&$1$&:&$0$&:&$0$&:&$0$   
\\
\iconSSK &   $0$&:&$0$&:&$1$&:&$-3$&:&$3$&:&$1$&:&$0$   
\\
\iconTwoSK &   $0$&:&$1$&:&$-4$&:&$6$&:&$-4$&:&$1$&:&$0$  
\\
\iconTwoSJ, \iconBSK &   $0$&:&$0$&:&$0$&:&$1$&:&$-2$&:&$1$&:&$0$   
\\
\hline
\hline
\end{tabular}
\end{table}
Considering the data taken at different temperatures, 
$\langle V_{ll'} V_{l'l}\rangle\propto \sigma_{xx}^{-1}-\sigma_{xx0}^{-1}$ approximately \cite{Hou2015PRL}, e.g., the resistivity $\sigma_{xx}^{-1}$ referenced from its lowest-temperature value $\sigma_{xx0}^{-1}$. Consequently, the third-order nonlinear Hall conductivity for the $\iconBSJ$ mechanism contains the $\sigma_{xx}^2$ and $\sigma_{xx}$ terms of opposite signs but comparable magnitudes (usually, $\sigma_{xx}\sim\sigma_{xx0}$ in the temperature range).
Moreover, the third-order skew-scattering mechanism (\iconThreeSK), i.e., three successive asymmetric scattering events off disorder, contributes to all seven parameters from $\mathcal{C}_{0}$ through $\mathcal{C}_{6}$ 
and has the weights
$\mathcal{C}_6 \sigma_{xx}^6
 :\mathcal{C}_5 \sigma_{xx}^5
 :\mathcal{C}_4 \sigma_{xx}^4
 :\mathcal{C}_3 \sigma_{xx}^3
 :\mathcal{C}_2 \sigma_{xx}^2
 :\mathcal{C}_1 \sigma_{xx} 
 :\mathcal{C}_0
 =  
 1:-6:15:-20:15:-6:1$, which is derived from an expansion of  $\sigma_{xx}^{6}(\sigma_{xx}^{-1}-\sigma_{xx0}^{-1})^6$ in the calculation of the scaling law, where the first power of 6 is from the $\tau^6$ in the distribution function $\delta f^{\iconThreeSK}$ and the second power of $6$ counts for three events of asymmetric scattering, each event contributes a power of 2.
In conclusion, the weight ratios of the polynomial terms in the scaling law encode characteristic features of different mechanisms, providing experimental fingerprints. See Sec.~\ref{Sec:calculation-scaling} for the derivation procedure of the scaling law.

\begin{figure*}[htbp]
\centering
\includegraphics[width=0.95\textwidth]{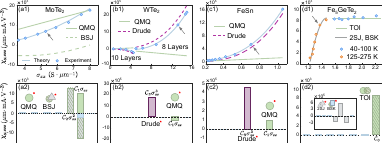}
\caption{The results of fitting the experimentally measured third-order nonlinear Hall conductivity $\chi_{y;xxx}$ as a function of the longitudinal conductivity $\sigma_{xx}$ (solid curves with circles), using the scaling laws in Eqs.~(\ref{Eq:Scaling-odd}) and (\ref{Eq:Scaling-even}). [(a1), (a2)] Results for MoTe$_2$ in the temperature range $T \in [10,100]~\mathrm{K}$ (adapted from Ref.~\cite{Lai21nn}). (a1) The green solid and dashed lines represent the quantum metric quadrupole (QMQ, \iconQMQ) mechanism and the mixed Berry-curvature-plus-side-jump (BSJ, \iconBSJ) mechanism, respectively. (a2) The corresponding scaling-law weights of the QMQ and BSJ contributions at $\sigma_{xx}=5.28~\mathrm{S}\cdot\mu\mathrm{m}^{-1}$.
[(b1), (b2)] Results for eight and ten layers of WTe$_2$ in the temperature range $T \in [2,80]~\mathrm{K}$ (adapted from Ref.~\cite{Ye22prb}). (b1) The green solid and purple dashed lines represent the QMQ mechanism and the Drude mechanism, respectively. (b2) The scaling-law weights of the Drude and QMQ contributions at $\sigma_{xx}=13.3~\mathrm{S}\cdot\mu\mathrm{m}^{-1}$.
[(c1), (c2)] The same analysis as in (b1) and (b2) but for FeSn in the temperature range $T \in [90,330]~\mathrm{K}$ (adapted from Ref.~\cite{Sankar24prx}).
[(d1), (d2)] The same as in (a1) and (a2) but for Fe$_5$GeTe$_2$ in the temperature range $T \in [40,275]~\mathrm{K}$ (adapted from Ref.~\cite{Yu25ncomms}). The dominant mechanism is the third-order intrinsic mechanism (TOI, \iconTOI) when $T < 100~\mathrm{K}$. For $T > 100~\mathrm{K}$, the response is jointly dominated by the second-order side-jump (2SJ, \iconTwoSJ) mechanism and the mixed Berry-curvature-plus-skew-scattering (BSK, \iconBSK) mechanism, with their scaling-law weights at $\sigma_{xx}=1.36~\mathrm{S}\cdot\mu\mathrm{m}^{-1}$ shown in the inset. The gray arrows indicate the values of $\sigma_{xx}$ used for extracting the weight ratios. The fitted scaling parameters and their confidence intervals for each experimental dataset are summarized in Sec.~SIV in Supplemental Material~\cite{supp}.}
\label{Fig:scalingWeight}
\end{figure*}

\section{Experimental applications}

\subsection{Scaling-Law Analysis}

We now apply the scaling law to analyze experimental data. 
In experiments, the third-order nonlinear Hall conductivity can be extracted using 
\begin{eqnarray}
\chi_{y;xxx}= \frac{V^{3\omega}_{y}\sigma_{xx}}{(V^{\omega}_{x})^{3}}\frac{L^3}{W},
\end{eqnarray}
where $L$ is the device length, $W$ is the width in 2D or cross-sectional area in 3D, $V^{3\omega}_{y}$ is the third-harmonic transverse voltage, and $V^{\omega}_{x}$ is the first-harmonic longitudinal voltage. Using Eqs.~(\ref{Eq:Scaling-odd}) and (\ref{Eq:Scaling-even}), we fit published datasets of $\chi_{y;xxx}$ versus $\sigma_{xx}$ for a variety of materials, including  FeTe \cite{Shao25arXiv}, Cd$_3$As$_2$ \cite{Tongyang23prl}, TaIrTe$_4$ \cite{Wang22nsr}, (SnS)$_{1.17}$(NbS$_2$)$_3$ \cite{Li24acsami}, MnBi$_2$Te$_4$ \cite{Li24ncomms}, FeSn \cite{Sankar24prx}, VSe$_2$ \cite{Chen24prb},  WTe$_2$ \cite{Ye22prb}, MoTe$_2$ \cite{Lai21nn}, Fe$_5$GeTe$_2$ \cite{Yu25ncomms}, and RuO$_2$ \cite{Chu25prl}.

For example, Figs.~\ref{Fig:scalingWeight}(a1) and~\ref{Fig:scalingWeight}(a2) show the fit for MoTe$_2$ \cite{Lai21nn}. The fit reveals that only the 
$\mathcal{C}_2\sigma_{xx}^2$ and $\mathcal{C}_1\sigma_{xx}$ terms are sizable, implying that the dominant contributions are either the quantum metric quadrupole mechanism (QMQ, \iconQMQ) or the mixed Berry-curvature-plus-side-jump mechanism (BSJ, \iconBSJ), according to Table~\ref{Tab:Weights}. Through confidence analysis (Sec.~SIV in Supplemental Material \cite{supp}), we find that the formula 
$\chi_{y;xxx}=\mathcal{C}^{\mathrm{QMQ}}\sigma_{xx}\sigma_{xx0}^{-1}+\mathcal{C}^{\mathrm{BSJ}}(\sigma_{xx}^2\sigma_{xx0}^{-2}-\sigma_{xx}\sigma_{xx0}^{-1})$
gives the best fit, with fitting parameters $\mathcal{C}^{\mathrm{BSJ}}\approx 2.15\times10^4$ and $\mathcal{C}^{\mathrm{QMQ}}\approx 1.75\times10^4$, both in units of $\mu \mathrm{m}\cdot \mathrm{mA} \cdot \mathrm{V}^{-3}$. Figure~\ref{Fig:scalingWeight}(a1) compares the QMQ (solid curve) with the BSJ (dashed curve). The QMQ term has the same sign as the total third-order nonlinear conductivity and is the dominant contribution in the experiment \cite{Lai21nn}.  Also, Fig.~\ref{Fig:scalingWeight} (a1) and later Fig.~\ref{Fig:scalingWeight} (d1) show that the nonlinear Hall conductivity is of the magnitude of 10$^3$ $\mu$m$\cdot$mA$\cdot$V$^{-3}$ when the geometric mechanisms are dominant.

Similarly, Figs.~\ref{Fig:scalingWeight}(b1) and~\ref{Fig:scalingWeight}(b2) show the fit for WTe$_2$ \cite{Ye22prb}. The formula
$\chi_{y;xxx}=\mathcal{C}^{\mathrm{Drude}}\sigma_{xx}^3\sigma_{xx0}^{-3}+\mathcal{C}^{\mathrm{QMQ}}\sigma_{xx}\sigma_{xx0}^{-1}$
gives the best fit, with fitting parameters $\mathcal{C}^{\mathrm{Drude}}\approx 9.60\times10^4$ and $\mathcal{C}^{\mathrm{QMQ}}\approx -1.37\times10^5$. The data of FeSn in Figs.~\ref{Fig:scalingWeight}(c1) and~\ref{Fig:scalingWeight}(c2) also show the same combination\textemdash the Drude mechanism and quantum metric quadrupole mechanism\textemdash as the dominant contributions, with fitting parameters $\mathcal{C}^{\mathrm{Drude}}\approx1.26\times10^6$ and $\mathcal{C}^{\mathrm{QMQ}}\approx 1.25\times10^5$. 
In both the WTe$_2$ and FeSn experiments, the Drude mechanism is stronger than the quantum metric quadrupole mechanism, although the quantum metric quadrupole mechanism is still sizable. Both experiments further indicate that, when the Drude mechanism dominates, the third-order nonlinear Hall conductivity is on the order of 10$^5$--10$^6$ $\mu$m$\cdot$mA$\cdot$V$^{-3}$.

Moreover, some experiments are dominated by disorder scattering, such as Fe$_5$GeTe$_2$ in Figs.~\ref{Fig:scalingWeight}(d1) and~\ref{Fig:scalingWeight}(d2). For $T > 100~\mathrm{K}$, the signal is jointly dominated by the second-order side-jump (2SJ, \iconTwoSJ) mechanism and the mixed Berry-curvature-plus-skew-scattering (BSK, \iconBSK) mechanism. The confidence analysis indicates that $\chi_{y;xxx}=\mathcal{C}_3\sigma^{3}_{xx}+\mathcal{C}_2\sigma^{2}_{xx}+\mathcal{C}_1\sigma_{xx}$ is the optimal fit. 
We further find that the three terms in the inset in Fig.~\ref{Fig:scalingWeight} (d2) have the ratios 
$\mathcal{C}_3 \sigma_{xx}^3:\mathcal{C}_2 \sigma_{xx}^2:\mathcal{C}_1 \sigma_{xx} = 1.00:-2.11:1.08$,
roughly consistent with the scaling-law weight ratios ($1:-2:1$) for the 2SJ and BSK mechanisms (row 13 in Table~\ref{Tab:Weights}), 
where the different signs and weights are the result of expanding $(\sigma^{-1}_{xx}-\sigma^{-1}_{xx0})^2$ in the derivation of the scaling law. Remarkably, upon cooling below 100 K, the nonlinear conductivity flattens and becomes almost independent of $\sigma_{xx}$, in sharp contrast to its behavior above 100 K. This indicates that, within the 40--100 K range, the nonlinear Hall response is dominated by the third-order intrinsic (TOI, \iconTOI) mechanism. 
Because the TOI mechanism requires broken time-reversal symmetry, the nonlinear Hall conductivity is consistent with the emergence of ferromagnetic order around 100 K in Fe$_5$GeTe$_2$~\cite{Yu25ncomms}.
TOI arises from a geometric coupling between the Berry curvature and the quantum metric \cite{Fang24prl} and is, therefore, an intrinsic, scattering-independent contribution from the quantum geometry of the Bloch bands. More importantly, the case of Fe$_5$GeTe$_2$ further shows that the nonlinear Hall effect is a powerful spectroscopic tool to study condensed matter phase transitions.

If the experimental results are better fitted by other functions,  such as exponential or sinusoidal functions, our scaling law in terms of polynomials is not applicable. Empirically, $\sigma_{xx}$ in the temperature window is not supposed to change by orders. The reliability of the fitting is supported by both the confidence interval and the coefficient of determination (see Sec.~SIV in Supplemental Material \cite{supp} for details). First, the confidence interval analysis shows that none of the parameters cross zero. Second, the fit yields the highest coefficient of determination (very close to 1). Therefore, we show that the geometric contributions to the third-order nonlinear conductivity, e.g., here, the quantum metric quadrupole and third-order intrinsic mechanism, can be identified in a disordered material using the scaling law derived in this work.

\subsection{Symmetry analysis}

Symmetry analysis can help reduce the polynomial terms in the scaling-law analysis. Because of the $\mathcal{T}$-odd mechanisms, the symmetry analysis is based on the magnetic point groups \cite{Litvin13Magnetic}. For the most common measurement setup on the $x$-$y$ plane, all the magnetic point-group symmetry operations can be described by eight elementary operations and their combinations. \begin{table}[thbp]
\caption[Symmetry of the third-order nonlinear Hall conductivity on the $x$-$y$ plane]{Symmetry of the third-order nonlinear Hall conductivities $\chi_{y;xxx}$ [Eq.~(\ref{Eq:definition-chi})] on the $x$-$y$ plane,  
under eight elementary symmetry operators, including the rotations about the $z$ axis $\{C_2^z, C_3^z, C_4^z, C_6^z\}$, the twofold rotation about $x$ or $y$ axis $\{ C_2^x, C_2^y\}$, spatial inversion $\mathcal{P}$, and time-reversal symmetry $\mathcal{T}$. They can combine to study all the magnetic point-group symmetry operations on the $x$-$y$ plane. 
The mechanisms are classified into four categories. The superscripts ``disorder" with and without the asterisk correspond to the disorder-related mechanisms that are even and odd under time reversal, respectively. 
Check mark (\Checkmark) means that the third-order nonlinear response is allowed by the symmetry. Sign $-$ and $+$ mean $\chi$ changes sign or not under the symmetry operation. $C_4^z$ relates the antisymmetric properties under the exchange of $x$ and $y$.}
\label{Tab:symmetry}
\setlength{\tabcolsep}{0.30cm}
\renewcommand\arraystretch{1.5}
\begin{tabular}{lccccc}
\hline
\hline
    &  $\mathcal{T}$    
&$C^{z}_{2}$, $\mathcal{P}$ & $C^{x,y}_{2}$   &$C^{z}_{3,6}$ 
 &$C^{z}_{4}$
\\ 
    \hline 
$\chi^{\mathrm{QMQ}}_{y;xxx}$   &  +   
&+ & $-$   &\Checkmark 
 &$-\chi_{x;yyy}$ 
\\
$\chi^{\mathrm{BCQ/TOI}}_{y;xxx}$ &  $-$  
&+ & $-$   &\Checkmark  
 &$-\chi_{x;yyy}$ 
\\
$\chi^{\mathrm{Disorder{\color{red}*}}}_{y;xxx}$  &  +   
&+ & $-$   &\Checkmark 
 &$-\chi_{x;yyy}$ 
\\
$\chi^{\mathrm{Disorder}}_{y;xxx}$  &  $-$  &+ & $-$   &\Checkmark   &$-\chi_{x;yyy}$ \\
\hline
\hline
\end{tabular}
\end{table}According to the behaviors of the third-order nonlinear conductivity $\chi$, the eight elementary operations can be classified into five cases and the mechanisms can be classified into four categories, as shown in Table \ref{Tab:symmetry}. 

(i) Under time-reversal symmetry $\mathcal{T}$, the $\mathcal{T}$-even mechanism remains invariant, such as the quantum metric quadrupole (\iconQMQ), while $\mathcal{T}$-odd mechanism changes sign, such as the Berry curvature quadrupole (\iconBCQ). 

(ii) Under $C_2^z$ and $\mathcal{P}$, all mechanisms remain invariant. Although they introduce negative signs to the $x$ and $y$ directions, the four $x$ or $y$ components in the Hall responses $\chi_{y;xxx}$ are equal to a positive sign.

(iii) Both of the in-plane twofold rotational symmetries $C^{x}_2$ and $C^{y}_2$ give a sign change of $\chi_{y;xxx}$ and $\chi_{x;yyy}$ for all mechanisms, because there are an odd number of $x$ or $y$ in the subscripts and the operations correspond to a sign change of $x$ or $y$. The sign change means that they alone forbid all mechanisms but can be combined with other operations to allow certain mechanisms.

(iv) $C_3^z$ and $C_6^z$ permit all mechanisms, while $\chi_{y;xxx}$ is not invariant but is related to other third-order tensor components of $\chi$.

(v) $C_4^z$ permits all mechanisms and imposes an antisymmetric relation between the two third-order Hall components 
\begin{equation}
 \chi_{y;xxx}=-\chi_{x;yyy}.   
\end{equation}

Now, we give three examples to show how the eight elementary symmetry operations in Table~\ref{Tab:symmetry} can be
combined to analyze magnetic point-group operations on the $x$-$y$ plane. 

(a) The mirror reflection about the $y-z$ plane $\sigma_x$ can be written as the combinations $\sigma_x=C_2^x\mathcal P$, which forbids
$\chi_{y;xxx}$ and $\chi_{x;yyy}$, because these components are odd $(-)$ under
$C_2^x$ but even $(+)$ under $\mathcal P$ and, hence, are odd $(-)$ under $\sigma_x$. 

(b) The antiunitary symmetry $C_3^z\mathcal T$ forbids the $\mathcal T$-odd mechanisms. The presence of $C_3^z\mathcal T$ implies
\(    (C_3^z\mathcal T)^3=\mathcal T ,\)
since $(C_3^z)^3=1$. Therefore, $\mathcal T$ itself is a symmetry of the
magnetic point group. According to Table~\ref{Tab:symmetry}, all
$\mathcal T$-odd mechanisms, such as BCQ or TOI and disorder, change sign under
$\mathcal T$ and must vanish, whereas only $\mathcal T$-even mechanisms are
allowed.  

(c) $C_6^z\mathcal T$ implies $(C_6^z\mathcal T)^3=C_2^z\mathcal T$. Since $\chi_{y;xxx}$ is even $(+)$ under $C_2^z$ but odd $(-)$ under $\mathcal T$ for $\mathcal T$-odd mechanisms, $C_2^z\mathcal T$ changes its sign and, hence, both $C_3^z\mathcal T$ and $C_6^z\mathcal T$ forbids the $\mathcal T$-odd mechanisms.

The second-order nonlinear Hall \cite{Du2019NC,Nandy2019PRB,Xiao2019PRB,Du2021NC,Ho2021NE,He2021NC,Watanabe2021PRR,Tiwari2021NC,Levchenko2021AP,Duan2022PRL,He2022NatNano,Huang22nsr,sinha22np,Kaplan23nc,Mao2023NC,Jin23prl,Yokouchi23prl,Huang23prl,Kaplan24prl,ChenR24prbl,Xiang24prb,ChenC24prr,YaoJ24prb,Wang24prl,makushko24ne,Lihm24prl} response and nonlinear longitudinal response \cite{Su24prl,Wang24prb} are forbidden by spatial inversion
$\mathcal{P}$ for all mechanisms, by $\mathcal{C}_3^z$ for all quantum geometry-related mechanisms, and by $\mathcal{T}$ for quantum metric-related mechanisms. In contrast, the third-order nonlinear response is allowed for these three symmetries and, therefore, greatly expands the range of material candidates for experimental and theoretical investigations.

\subsection{Protocol of data analysis}

Finally, we summarize the protocol of using the scaling law and symmetry to analyze experimental data.

The first step is a symmetry diagnosis, to determine whether time-reversal
symmetry $\mathcal{T}$ is present and to identify the magnetic point-group symmetry of the device. The symmetry constraints then specify the allowed nonlinear conductivity components and eliminate symmetry-forbidden mechanisms.

The second step is to select the scaling law. Systems without and with time-reversal symmetry should be analyzed with Eqs.~(\ref{Eq:Scaling-odd}) and (\ref{Eq:Scaling-even}), respectively. 

The third step is to fit the experimental data of $\chi_{y;xxx}$ as a function of $\sigma_{xx}$, e.g., in a temperature range, to extract the scaling parameters $\mathcal{C}_n$. A confidence analysis is needed to keep the statistically significant $\mathcal{C}_n$. 
If a $\mathcal{C}_n$ has a confidence interval including zero, e.g., 0.1 $\pm$ 0.2, it is not statistically distinguishable from zero and should be abandoned. 

The final step is to identify mechanisms, by matching the fitted $\mathcal{C}_n$ with Table~\ref{Tab:Weights}. The set of $\mathcal{C}_n$ may correspond to multiple mechanisms, so a mechanism-resolved $\chi_{y;xxx}-\sigma_{xx}$ scaling law can be proposed to transform $\mathcal{C}_n$ into mechanism parameters, e.g., $\mathcal{C}^\mathrm{Drude}$, $\mathcal{C}^\mathrm{QMQ}$, and $\mathcal{C}^\mathrm{BSJ}$. The mechanism parameters then give the weights of all possible mechanisms as those in Fig.~\ref{Fig:scalingWeight}.

\section{Methods and Calculations}
\subsection{Calculation of third-order nonlinear conductivity}\label{Sec:calculation-conductivity}

The nonlinear conductivity $\chi_{a;bcd}$ can be calculated from the electric current density 
$\mathbf{J}=-e \sum_{l} \dot{\mathbf{r}}_{l} f_{l}$ in response to an electric field $\mathbf{E}= \left( \mathcal{E} e^{i\omega t}+\mathcal{E} ^*e^{-i\omega t} \right)/2$ of amplitude $\mathcal{E}$ and frequency $\omega$, where $-e$ is the electron charge and $l$ labels the band index $\gamma$ and wave vector $\mathbf{k}$. The velocity has five terms
\begin{eqnarray}\label{eq:velocity}
\dot{\mathbf{r}}_{l}=\boldsymbol{v}_{l}-\dot{\mathbf{k}} \times\left( \boldsymbol{\Omega}_{l}+ \nabla_\mathbf{k}\times\boldsymbol{\mathcal{G}}\mathbf{E}\right)+\boldsymbol{v}_{l}^{sj}+\boldsymbol{v}_{l}^{\mathrm{TOI}},
\end{eqnarray}
where the group velocity $\boldsymbol{v}_l=(1/\hbar)(\partial \epsilon_k^\gamma/\partial \mathbf{k})$ with the energy $\epsilon_k^\gamma$. In the absence of a magnetic field, $\dot{\mathbf{k}}=-(e/\hbar) \mathbf{E}$, where $\hbar$ is the reduced Planck constant. The Berry curvature~\cite{Xiao10rmp} 
\begin{eqnarray}
\boldsymbol{\Omega}^{a}_{l}=-2\varepsilon^{abc}\sum_{\gamma^{\prime}\ne \gamma} \mathrm{Im}\langle\gamma|\partial^{b}_{\mathbf{k}}\hat{\mathcal{H}}|\gamma^\prime\rangle\langle\gamma^\prime|\partial^{c}_{\mathbf{k}}\hat{\mathcal{H}}|\gamma\rangle/(\epsilon^{\gamma}_{\mathbf{k}}-\epsilon^{\gamma^\prime}_{\mathbf{k}})^2, \notag \\
\end{eqnarray}
where $\varepsilon^{abc}$ is the Levi-Civita symbol, $a,b,c\in \{x,y,z\}$,  $\partial^{a}_{\mathbf{k}}\equiv \partial/\partial k_a$, $\nabla_\mathbf{k}\equiv (\partial^x_{\mathbf{k}},\partial^y_{\mathbf{k}},\partial^z_{\mathbf{k}})$, and $\hat{\mathcal{H}}$ is the Hamiltonian. The Berry connection polarizability tensor~\cite{GaoY14prl}  
\begin{eqnarray}
\boldsymbol{\mathcal{G}}^{ab}_{l}=2e\sum_{\gamma^{\prime}\ne \gamma}\mathrm{Re}A^{\gamma\gamma^\prime}_{a}A^{\gamma^\prime\gamma}_{b}/(\epsilon^{\gamma}_{\mathbf{k}}-\epsilon^{\gamma^\prime}_{\mathbf{k}}),
\end{eqnarray} where the Berry connection $A^{\gamma\gamma^\prime}_{a}=i\langle\gamma|\partial^a_{\mathbf{k}}|\gamma\rangle$. $\boldsymbol{v}_{l}^{sj}$ is the side-jump velocity \cite{Nagaosa10rmp}. 
The second-order electric-field correction to the anomalous velocity $\boldsymbol{v}_{l}^{\mathrm{TOI}}$ has various forms in the literature \cite{Fang24prl,Dimi23prb,WangJian23prb}. None of them depend on scattering, so they all lead to the same scaling law. The distribution function $f_l$ is found by solving the Boltzmann equation \cite{Du2019NC,Sinitsyn2008JPCM}
\begin{eqnarray}
\partial f_{l}/\partial t+\dot{\mathbf{k}} \cdot  \partial f_{l}/\partial \mathbf{k}=\mathcal{I}_{e l}\{f_{l}\},
\label{eq:boltzmann}
\end{eqnarray}
where $\mathcal{I} _{el}\{ f_{l} \} $ accounts for disorder scattering, including the side jump, skew scattering, processes described by the scattering time $\tau$, and their higher-order and mixed effects. Both $\dot{\mathbf{r}}_{l}$ and $f_l$ can be expanded as polynomials in $\mathbf{E}$. By matching the product $\dot{\mathbf{r}}_{l} f_l$ at the cubic order of $\mathbf{E}$,
we can find the third-order nonlinear conductivity $\chi_{a;bcd}$, defined as 
$J_a \equiv \chi_{a;bcd}E_b E_c E_d$, where $a,b,c,d\in \{x,y,z\}$ and $J_a$ and $E_{b,c,d}$ are components of $\mathbf{J}$ and $\mathbf{E}$, respectively. For example, the  nonlinear conductivity of the third-order Drude mechanism is found as
\begin{equation}\label{Eq:Drude}
\chi^{\mathrm{Drude}}_{a;bcd}=-\frac{e^4\tau^3}{\hbar^3}\sum_l v_l^a\partial _{\mathbf{k}}^{b}\partial _{\mathbf{k}}^{c}\partial _{\mathbf{k}}^{d}f_{l}^{(0)},
\end{equation}
where the Fermi distribution $f_l^{(0)}=1/[1+e^{(\epsilon_l-\epsilon_\mathrm{F})/k_\mathrm{B} T}]$ with the Fermi energy $\epsilon_\mathrm{F}$.
For completeness, we also list the third-order nonlinear conductivities of the geometric mechanisms, including the Berry curvature quadrupole (BCQ, \iconBCQ) \cite{KTLow23prb}, the quantum metric quadrupole (QMQ, \iconQMQ) \cite{Yang22prb}, and the third-order intrinsic (TOI, \iconTOI) mechanism which mixes quantum metric and Berry curvature \cite{Fang24prl}:
\begin{eqnarray}\label{Eq:QG-formula} 
\chi _{a;bcd}^{\mathrm{BCQ}}&=&\frac{e^4\tau ^2}{\hbar ^3}\sum_{l}{\varepsilon ^{amb}\partial _{\mathbf{k}}^{c}\partial _{\mathbf{k}}^{d}\Omega_l^{m}f_l^{(0)}}, \notag \\
\chi _{a;bcd}^{\mathrm{QMQ}}&=&\frac{e^4\tau}{\hbar ^2}\sum_{l}{\left( \partial _{\mathbf{k}}^{d}\partial _{\mathbf{k}}^{a}G_l^{bc}-\partial _{\mathbf{k}}^{d}\partial _{\mathbf{k}}^{b}G_l^{ac} \right)}f_l^{(0)},\notag \\
\chi _{y;xxx}^{\mathrm{TOI}}&=&\frac{2e^4}{\hbar}\sum_{l}G^{xx}_{l}\Omega^{z}_{l}f_l^{(0)}/(\epsilon_{\bar{l}}-\epsilon_{l}).
\end{eqnarray} 
More details on the calculation of third-order nonlinear conductivity can be found in Sec.~SI in Supplemental Material \cite{supp}.

\subsection{Calculation of the scaling law}\label{Sec:calculation-scaling}

We derive the scaling law, namely, the third-order nonlinear Hall conductivity $\chi_{y;xxx}$ as a polynomial function of the linear longitudinal conductivity $\sigma_{xx}$, by exploiting the fact that both of them are governed by disorder scattering. For example, the linear longitudinal conductivity $\sigma_{xx}$ is linearly proportional to the scattering time $\tau$ in the semiclassical regime \cite{Mahan1990}, i.e., 
\begin{eqnarray}
    \sigma_{xx} \propto \tau,
\end{eqnarray}
and, according to Eqs.~(\ref{Eq:Drude}) and (\ref{Eq:QG-formula}), the third-order nonlinear Hall conductivities of the Drude and three geometric (third-order intrinsic \iconTOI, quantum metric quadrupole \iconQMQ, and Berry curvature quadrupole \iconBCQ) mechanisms scale with $\tau$ as
\begin{eqnarray}
\iconTOI \propto\tau^0,\  \ \iconQMQ \propto\tau^1, \ \ \iconBCQ \propto\tau^2,  \ \ \mathrm{Drude} \propto \tau^3,  
\end{eqnarray}
respectively, so there are relations
\begin{eqnarray}
   \iconTOI \propto \sigma_{xx}^0,\  
\iconQMQ \propto \sigma_{xx}^1,\ 
\iconBCQ \propto \sigma_{xx}^2,  \  \mathrm{Drude} \propto \sigma_{xx}^3  . 
\end{eqnarray}

Moreover, the disorder-related mechanisms are also related to the 
correlations of scattering matrix elements $V_{ll'}$ between state $l$ and $l'$:

\begin{eqnarray}
   \langle V_{ll'} V_{l'l}\rangle  \sim  n_i V_0^2 &\sim &\sigma_{xx}^{-1}-\sigma_{xx0}^{-1}, \notag\\
 \langle V_{ll^{\prime\prime}}V_{l^{\prime\prime}l^\prime}V_{l^\prime l}\rangle \sim n_{i}V^{3}_1 &\sim &\sigma_{xx}^{-1}-\sigma_{xx0}^{-1}, \notag \\ \langle  V_{ll^{\prime \prime}}V_{l^{\prime\prime} l^{\prime}}\rangle \langle V_{ll^{\prime\prime\prime}}V_{l^{\prime\prime\prime} l^{\prime}}\rangle \sim n_{i}^2V^{4}_0 &\sim &(\sigma_{xx}^{-1}-\sigma_{xx0}^{-1})^2, \label{Eq:correlations}
\end{eqnarray}
where $\langle \cdots \rangle$ means the ensemble average over disorder configurations, $n_i$ is the impurity concentration, and $V_0$ and $V_1$ measure the correlations of two and three scattering events, respectively. 
$\sigma_{xx}^{-1}$ is the resistivity, and $\sigma_{xx0}^{-1}$ is the resistivity at the lowest temperature. The scattering correlations are related to the difference $\rho_{xx}-\rho_{xx0}$, because in practice the scaling-law analysis is based on the data at different temperatures \cite{Tongyang23prl,Wang22nsr,Li24acsami,Li24ncomms,Sankar24prx,Chen24prb,Yu25ncomms,Ye22prb,Lai21nn}.

By applying the rules in Eq.~(\ref{Eq:correlations}) to Fig.~\ref{Fig:Mechanism}, we can find the dependence of all mechanisms on $\tau$ and the correlations of scattering 
\begin{eqnarray}
\iconSK &\propto & \tau^3  \langle  V_{ll^{\prime \prime}}V_{l^{\prime\prime} l^{\prime}}\rangle \langle V_{ll^{\prime\prime\prime}}V_{l^{\prime\prime\prime} l^{\prime}}\rangle,   \notag \\
\iconTwoSK &\propto & \tau^4  (\langle  V_{ll^{\prime \prime}}V_{l^{\prime\prime} l^{\prime}}\rangle \langle V_{ll^{\prime\prime\prime}}V_{l^{\prime\prime\prime} l^{\prime}}\rangle)^2, \notag  \\
\iconThreeSK &\propto & \tau^6  (\langle  V_{ll^{\prime \prime}}V_{l^{\prime\prime} l^{\prime}}\rangle \langle V_{ll^{\prime\prime\prime}}V_{l^{\prime\prime\prime} l^{\prime}}\rangle)^3, \notag  \\
\iconSJ &\propto &\tau^3 \langle V_{ll'} V_{l'l}\rangle,  \  \ 
\iconQSJ \propto  \tau \langle V_{ll'} V_{l'l}\rangle, \notag  \\
\iconSSK &\propto & \tau^4 \langle V_{ll'} V_{l'l}\rangle(\langle  V_{ll^{\prime \prime}}V_{l^{\prime\prime} l^{\prime}}\rangle \langle V_{ll^{\prime\prime\prime}}V_{l^{\prime\prime\prime} l^{\prime}}\rangle),  \notag\\
\iconBSJ &\propto & \tau^2 \langle V_{ll'} V_{l'l}\rangle,   \  \
\mathrm{Drude} \propto  \tau^3,\notag \\
\iconTwoSJ &\propto &\tau^3(\langle V_{ll'} V_{l'l}\rangle)^2,\ \
\iconBSK  \propto  \tau^3(\langle  V_{ll^{\prime \prime}}V_{l^{\prime\prime} l^{\prime}}\rangle \langle V_{ll^{\prime\prime\prime}}V_{l^{\prime\prime\prime} l^{\prime}}\rangle),\notag\\
\iconTwoSSK &\propto& \tau^4(\langle  V_{ll^{\prime \prime}}V_{l^{\prime\prime} l^{\prime}}\rangle)^2(\langle  V_{ll^{\prime \prime}}V_{l^{\prime\prime} l^{\prime}}\rangle \langle V_{ll^{\prime\prime\prime}}V_{l^{\prime\prime\prime} l^{\prime}}\rangle),\notag\\
\iconBTwoSK &\propto& \tau^4(\langle  V_{ll^{\prime \prime}}V_{l^{\prime\prime} l^{\prime}}\rangle \langle V_{ll^{\prime\prime\prime}}V_{l^{\prime\prime\prime} l^{\prime}}\rangle)^2,\notag\\
\iconBTwoSJ &\propto& \tau^2( \langle  V_{ll^{\prime \prime}}V_{l^{\prime\prime} l^{\prime}}\rangle)^2,\ \ 
\iconThreeSJ  \propto  \tau^3( \langle  V_{ll^{\prime \prime}}V_{l^{\prime\prime} l^{\prime}}\rangle)^3,\notag \\
\iconQSK &\propto& \tau^2( \langle  V_{ll^{\prime \prime}}V_{l^{\prime\prime} l^{\prime}}\rangle \langle V_{ll^{\prime\prime\prime}}V_{l^{\prime\prime\prime} l^{\prime}}\rangle),\notag\\
\iconBSSK &\propto& \tau^3( \langle  V_{ll^{\prime \prime}}V_{l^{\prime\prime} l^{\prime}}\rangle)( \langle  V_{ll^{\prime \prime}}V_{l^{\prime\prime} l^{\prime}}\rangle \langle V_{ll^{\prime\prime\prime}}V_{l^{\prime\prime\prime} l^{\prime}}\rangle), \notag \\
\iconTOI &\propto &\tau^0,\ 
\iconQMQ \propto \tau^1,\ 
\iconBCQ \propto \tau^2,
\end{eqnarray}
where the scattering time $\tau$ is from the distribution functions $\delta f$. Because the scaling-law analysis is performed by varying temperature, the three-event correlation $\langle V_{ll^{\prime\prime}}V_{l^{\prime\prime}l^\prime}V_{l^\prime l}\rangle $ does not appear explicitly.
Using Eq.~(\ref{Eq:correlations}) and the relations between resistivity and conductivity  $\rho_{xx}=\sigma_{xx}^{-1}$ and $\rho_{xx0}=\sigma_{xx0}^{-1}$, we have 
\begin{eqnarray}
\iconSK &\propto & \sigma_{xx}^3  (\sigma_{xx}^{-1}-\sigma_{xx0}^{-1})^{2}, \ \
\iconTwoSK \propto  \sigma_{xx}^4  (\sigma_{xx}^{-1}-\sigma_{xx0}^{-1})^{4}, \notag  \\
\iconThreeSK &\propto & \sigma_{xx}^6  (\sigma_{xx}^{-1}-\sigma_{xx0}^{-1})^{6}, \ \
\iconSJ \propto \sigma_{xx}^3  (\sigma_{xx}^{-1}-\sigma_{xx0}^{-1}),   \notag\\
\iconQSJ &\propto & \sigma_{xx}  (\sigma_{xx}^{-1}-\sigma_{xx0}^{-1}),   \ \
\iconSSK \propto  \sigma_{xx}^4  (\sigma_{xx}^{-1}-\sigma_{xx0}^{-1})^{3}, \notag \\
\iconBSJ &\propto & \sigma_{xx}^2  (\sigma_{xx}^{-1}-\sigma_{xx0}^{-1}),   \ \
\mathrm{Drude} \propto \sigma_{xx}^3  , \notag\\
\iconTwoSJ &\propto& \sigma_{xx}^3  (\sigma_{xx}^{-1}-\sigma_{xx0}^{-1})^{2},\ \ 
\iconBSK \propto  \sigma_{xx}^3  (\sigma_{xx}^{-1}-\sigma_{xx0}^{-1})^{2},\notag\\
\iconTwoSSK &\propto& \sigma_{xx}^4  (\sigma_{xx}^{-1}-\sigma_{xx0}^{-1})^{4},\ \
\iconBTwoSK \propto \sigma_{xx}^4  (\sigma_{xx}^{-1}-\sigma_{xx0}^{-1})^{4},\notag\\
\iconBTwoSJ &\propto& \sigma_{xx}^2  (\sigma_{xx}^{-1}-\sigma_{xx0}^{-1})^{2},\ \
\iconQSK  \propto  \sigma_{xx}^2  (\sigma_{xx}^{-1}-\sigma_{xx0}^{-1})^{2},\notag\\
\iconBSSK &\propto& \sigma_{xx}^3  (\sigma_{xx}^{-1}-\sigma_{xx0}^{-1})^{3},\ \
\iconThreeSJ \propto \sigma_{xx}^3  (\sigma_{xx}^{-1}-\sigma_{xx0}^{-1})^{3},\notag \\ 
\iconTOI &\propto &\sigma_{xx}^0,\ 
\iconQMQ \propto \sigma_{xx}^1,\ 
\iconBCQ \propto \sigma_{xx}^2.
\end{eqnarray}
Usually, $\sigma_{xx}$ varies little over the measured temperature range, so  $\sigma_{xx}\sigma_{xx0}^{-1}\approx 1$. Under this approximation, we obtain the weights of the polynomial terms in the scaling law, as listed in Table~\ref{Tab:Weights}. More details on the derivation of the scaling law can be found in Sec.~SIII in Supplemental Material \cite{supp}.

\begin{acknowledgements}
This work is supported by the National Key R$\&$D Program of China (2022YFA1403700), 
Quantum Science and Technology-National Science and Technology Major Project (2021ZD0302400), the National Natural Science Foundation of China (12525401 and 12350402), Guangdong Basic and Applied Basic Research Foundation (2023B0303000011), Guangdong Provincial Quantum Science Strategic Initiative (GDZX2201001 and GDZX2401001), the Science, Technology and Innovation Commission of Shenzhen Municipality (ZDSYS20190902092905285),
High-level Special Funds (G03050K004), the New Cornerstone Science Foundation through the XPLORER PRIZE, and the Center for Computational Science and Engineering of SUSTech.
\end{acknowledgements}

\section*{DATA AVAILABILITY}
The data that are fitted in this article are available in Refs.~\cite{Lai21nn,Yu25ncomms,Ye22prb,Sankar24prx}.

\nocite{Tian2009PRL,Gong24arXiv}
\bibliographystyle{apsrev4-1-etal-title_6authors}
\bibliography{refs-NHE-PT-APS}

\end{document}